\pdfoutput=1

\documentclass[twocolumn,showpacs,amssymb,aps,nofootinbib,floatfix,superscriptaddress]{revtex4-1}

\usepackage{epsfig}
\usepackage{float}
\usepackage{amssymb}
\usepackage{hyperref}
\usepackage{amsmath}

\newcommand{\be}{\begin{equation}}
\newcommand{\ee}{\end{equation}}

\newcommand{\bea}{\begin{eqnarray}}
\newcommand{\eea}{\end{eqnarray}}
\newcommand{\ba}{\begin{array}}
	\newcommand{\ea}{\end{array}}

\begin{document}

\title{Factorization breaking for higher moments of harmonic flow}

\author{Piotr Bożek}
\email{piotr.bozek@fis.agh.edu.pl}
\affiliation{AGH University of Science and Technology, Faculty of Physics and
Applied Computer Science, aleja Mickiewicza 30, 30-059 Cracow, Poland}

\author{Rupam Samanta}
\email{rsamanta@agh.edu.pl}
\affiliation{AGH University of Science and Technology, Faculty of Physics and
Applied Computer Science, aleja Mickiewicza 30, 30-059 Cracow, Poland}

\begin{abstract}
We study correlations between  harmonic flow vectors squared measured at different transverse momenta. One of the flow harmonics squared is taken at a fixed transverse
momentum and correlated to the momentum averaged harmonic flow squared of the same order.
Such  four particle correlators, dependent on transverse momentum,
have been recently measured experimentally. Factorization coefficients based on the ratio of such four-particle correlators allow the independent measurement of the  flow vector and flow magnitude factorization breaking coefficient. Moreover, the correlation of the angles of flow harmonics  as a function of transverse momentum  can be extracted. Results are compared to  preliminary data of the ALICE Collaboration.
We also present  predictions for the momentum dependent factorization breaking coefficient between mixed flow harmonics. The correlators with squares of  mixed harmonics can serve as a way to independently measure
 the flow vector, flow magnitude, and flow angle correlations, and could be used to gain additional information on the fluctuating initial state and the dynamics in heavy-ion collisions. 
\end{abstract}


\maketitle

\section{Introduction}

One of the main goals of the experimental program in high-energy nuclear collisions is to study the properties of the dense matter created in the collision.
The dense matter created in the interaction region of the collision expands
and a collective flow of matter appears \cite{Ollitrault:2010tn,Heinz:2013th,Gale:2013da}. One of the essential characteristics of high-energy collisions is the
presence of event-by-event fluctuations in the initial state \cite{Aguiar:2001ac,Takahashi:2009na,PHOBOS:2006dbo,Alver:2010gr,Schenke:2010rr,Schenke:2012wb,Teaney:2010vd}. Initial state fluctuations manifest themselves in 
event by event fluctuations of the collective flow generated in the collisions.

One of the aspects  of collective flow fluctuations is the decorrelation
of the harmonic flow vectors  measured   at different momenta.
In the longitudinal direction, it means that the correlation coefficient
for the harmonic flow vectors measured at two different pseudorapidities
is smaller than one \cite{Bozek:2010vz,Jia:2014ysa,Pang:2014pxa,Pang:2015zrq,CMS:2015xmx,Bozek:2017qir,Cimerman:2021gwf}.
In the transverse momentum
it manifests itself as a deviation from one of the correlation coefficient
 (called in this context the factorization breaking coefficient)
 for flow vectors measured at two different transverse momenta
\cite{Gardim:2012im,Kozlov:2014fqa,Gardim:2017ruc,Zhao:2017yhj,Bozek:2018nne,Barbosa:2021ccw}.

The decorrelation  between two flow vectors is due to the decorrelation of the flow vector magnitudes and of the  flow vector angles
\cite{Heinz:2013bua,Jia:2014ysa}. The two effects can be separated
when four-particle correlators
in  pseudorapidity are measured \cite{Jia:2017kdq,ATLAS:2017rij}. For
the correlation of flow vectors in transverse momentum a similar procedure
would involve the measurement of four-particle correlators for harmonic
flow vectors at different transverse momenta \cite{Bozek:2018nne}.
Such four-particle correlators have been  measured experimentally
only for the case when two harmonic flow vectors  are defined
at fixed transverse momentum and two are momentum averaged \cite{NielsenIS2021}.

We present calculations for the flow vectors squared  and flow magnitudes squared  factorization breaking coefficients and compare them to the preliminary results of the ALICE Collaboration. We check in the model that the estimation of the flow angle decorrelation used by the ALICE Collaboration is a good measure of the event by event
flow angle decorrelation between harmonic flow vectors. The
flow angle decorrelation between first moments of
flow vectors cannot be measured,
but can be approximated as the square root of the flow angle decorrelation
between two  flow  vectors squared. Finally, we present prediction for the momentum dependent correlation between mixed harmonics. Also in that case, using squares of flow vectors (or flow magnitudes) makes possible the separate  measurement of  the flow vector, flow vector magnitude, and flow angle correlations between mixed harmonics.

\section{Factorization breaking for harmonic flow vectors and flow  magnitudes}


The harmonic flow coefficients measure the azimuthal asymmetry in the distribution of particles emitted in a collision. The particle distribution is written as
\begin{equation}
\frac{dN}{dp d\phi}=\frac{dN}{2\pi dp}\left( 1+  2 \sum_{n=1}^{\infty} V_n(p) e^{i n\phi}\right) \ .
\end{equation}
The complex vector $V_n(p)=v_n(p)  e^{i n \Psi_n(p)}$ denotes the harmonic flow vector of order $n$ for particles emitted at the transverse momentum $p$; $v_n$ and $\Psi_n$ are the corresponding flow magnitude and angle.
The harmonic flow averaged over the whole  range of transverse momentum is  $V_n$.
The average over a  sample of events in a given centrality bin is denoted by $\langle \dots \rangle$. The two-particle cumulant  formula for the harmonic flow coefficient  takes the form 
\begin{equation}
v_n\{2\}= \sqrt{ \langle V_n V_n^\star \rangle}
\end{equation}
for the momentum averaged  flow and
\begin{equation}
v_n\{2\}(p)= \frac{ \langle V_n V_n^\star(p) \rangle}{\sqrt{\langle V_n V_n^\star\rangle}}
\label{eq:v22}
\end{equation}
for the momentum dependent flow.

In this paper we study the harmonic flow  as a function of transverse momentum. All results presented are obtained using a boost invariant version of the hydrodynamic model MUSIC  \cite{Schenke:2010nt,Schenke:2010rr,Paquet:2015lta} for Pb+Pb collisions at $5.02$ TeV.
The initial entropy distributions  for the hydrodynamic evolution are taken from two models;  a two-component Glauber Monte Carlo model \cite{Bozek:2019wyr} and the TRENTO model \cite{Moreland:2014oya}.
 Unless otherwise stated, we use a constant shear viscosity to entropy density ratio  $\eta/s=0.08$.

Event by event fluctuations in the initial conditions
cause fluctuations in the final harmonic flow. One of the effects of these fluctuations  is the   decorrelation  between the harmonic flow vectors in two different kinematic regions  \cite{Bozek:2010vz,Gardim:2012im}. The correlation between flow vectors at two different transverse momenta can be measured using the factorization breaking coefficient (correlation coefficient)
\begin{equation}
r_n(p_1,p_2)=\frac{\langle V_n(p_1) V_n^\star(p_2) \rangle}{\sqrt{\langle v_n^2(p_1) \rangle \langle v_n^2(p_2) \rangle}} \ .
\label{eq:facb}
\end{equation}
The factorization breaking  coefficient $r_n(p_1,p_2)$ has been measured in experiment
\cite{CMS:2015xmx,CMS:2013bza,Zhou:2014bba} and calculated
in models
\cite{Gardim:2012im,Kozlov:2014fqa,Gardim:2017ruc,Zhao:2017yhj,Bozek:2018nne,Barbosa:2021ccw}.

The factorization breaking coefficient  (\ref{eq:facb}) is defined as the
correlation coefficient between two flow vectors (as complex numbers).
In the following, we study similar factorization breaking
 coefficients between flow vectors squared and flow vector
magnitudes squared, as well as for mixed flow harmonics. 
In this context, the decorrelation is understood as the deviation of
these factorization breaking coefficient from one. The quantity defined as the correlation for the flow
angles between two flow vectors involves the cosine of the angle difference, which
is not the Pearson correlation coefficient. It is  the circular correlation coefficient weighted with flow vector magnitudes.  In that case we use the names correlation between
flow vector angles, and flow angle decorrelation.

The decorrelation  between harmonic flow vectors at two different
momenta involves both the decorrelation of the flow magnitudes and of the flow angles
\cite{Heinz:2013bua,Jia:2014ysa}. The flow magnitude or the flow angle decorrelations  cannot be measured experimentally with two-particle correlators. The flow angle or flow magnitude decorrelation in pseudorapidity
can be estimated independently of the flow vector decorrelation using
four-particle correlators \cite{Jia:2017kdq}.  Previous experimental results of the ATLAS Collaboration have shown that
the decorrelation of harmonic flow vectors
in pseudorapidity is composed in roughly equal strength from flow magnitude
and flow angle decorrelations \cite{ATLAS:2017rij}.

For the correlation  of harmonic flow vectors at two different
transverse momenta a similar procedure can be used \cite{Bozek:2018nne}.
Four-particle correlators, measuring the factorization breaking between
two flow vectors {\it squared}, can be defined as,
\begin{equation}
r_{n;2}(p_1,p_2)=\frac{\langle V_n^2(p_1) V_n^{\star\ }(p_2)^{2} \rangle}{\sqrt{\langle v_n^4(p_1)  \rangle \langle v_n^{ 4}(p_2) \rangle}}
\end{equation}
and for the flow magnitudes {\it squared}, it is constructed as,
\begin{equation}
\begin{aligned}
r_{n}^{v_n^2}(p_1,p_2)=\frac{\langle |V_n(p_1)|^2 |V_n(p_2)|^2 \rangle}{\sqrt{\langle
v_n^4(p_1)  \rangle \langle v_n^{ 4}(p_2) \rangle}}\\ = \frac{\langle v_n^2(p_1) v_n^{ 2}(p_2) \rangle}{\sqrt{\langle v_n^4(p_1)  \rangle \langle v_n^{ 4}(p_2) \rangle}}
\end{aligned}
\end{equation}
In practice, the above formulae are difficult to use
in experiment due to low statistics in high momentum bins.

\section{Results for Pb+Pb collisions at $\sqrt{s_{NN}}=5.02$~TeV}

\label{sect:results}

\begin{figure}
\vspace{5mm}
\begin{center}
\includegraphics[width=0.48 \textwidth]{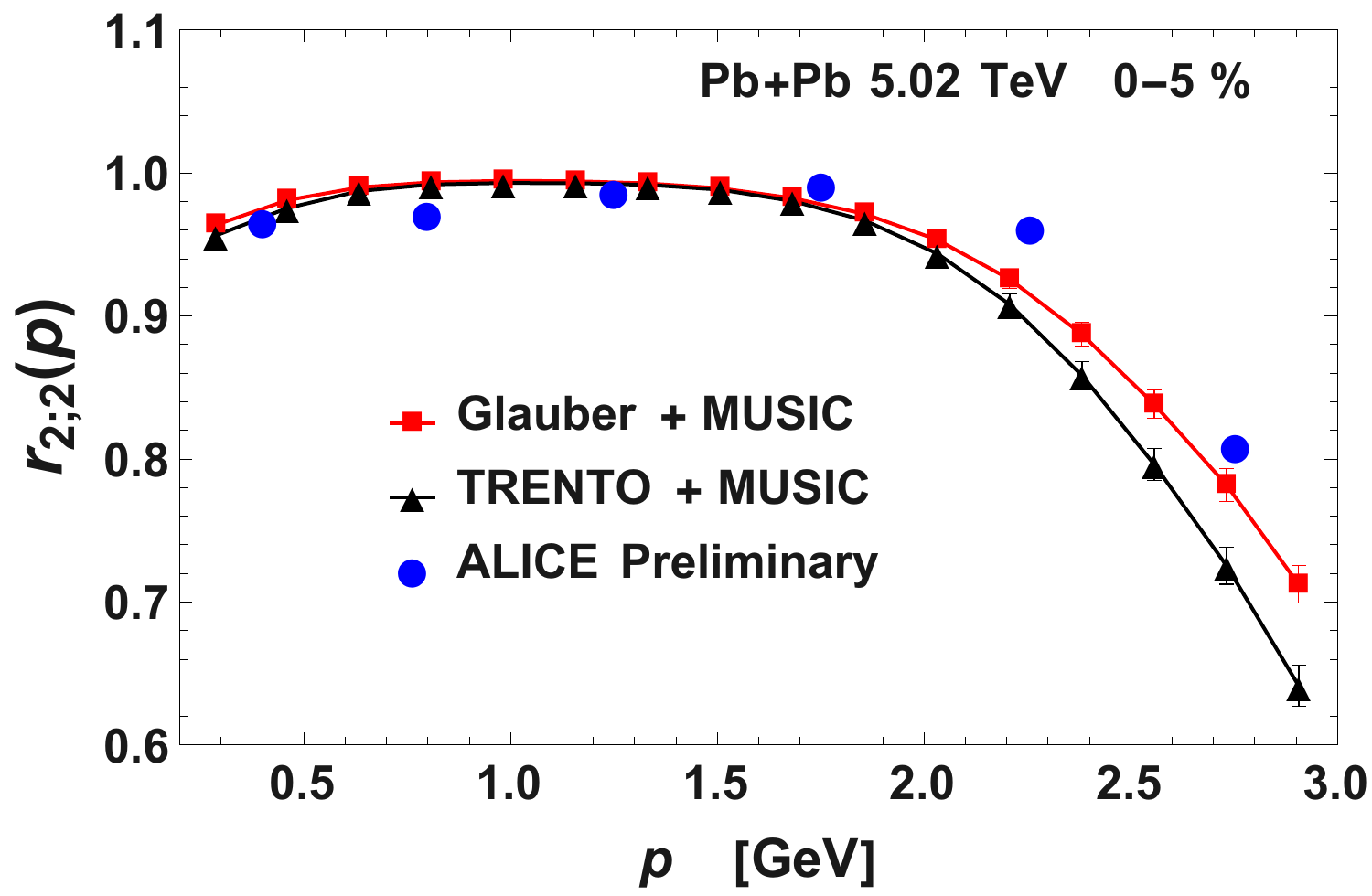} 
\end{center}
\caption{The factorization breaking coefficient between flow vectors squared $V_2(p)^2$ and $V_2^2$  as a
function of the transverse momentum $p$,
 for Pb+Pb collisions with centrality $0-5$\%. The red squares and black triangles denote the results obtained with the Glauber model and the TRENTO model initial conditions respectively. The experimental data of the ALICE Collaboration are represented using  blue dots \cite{NielsenIS2021}.}
\label{fig:r2205}
\end{figure}

\begin{figure}
\vspace{5mm}
\begin{center}
\includegraphics[width=0.48 \textwidth]{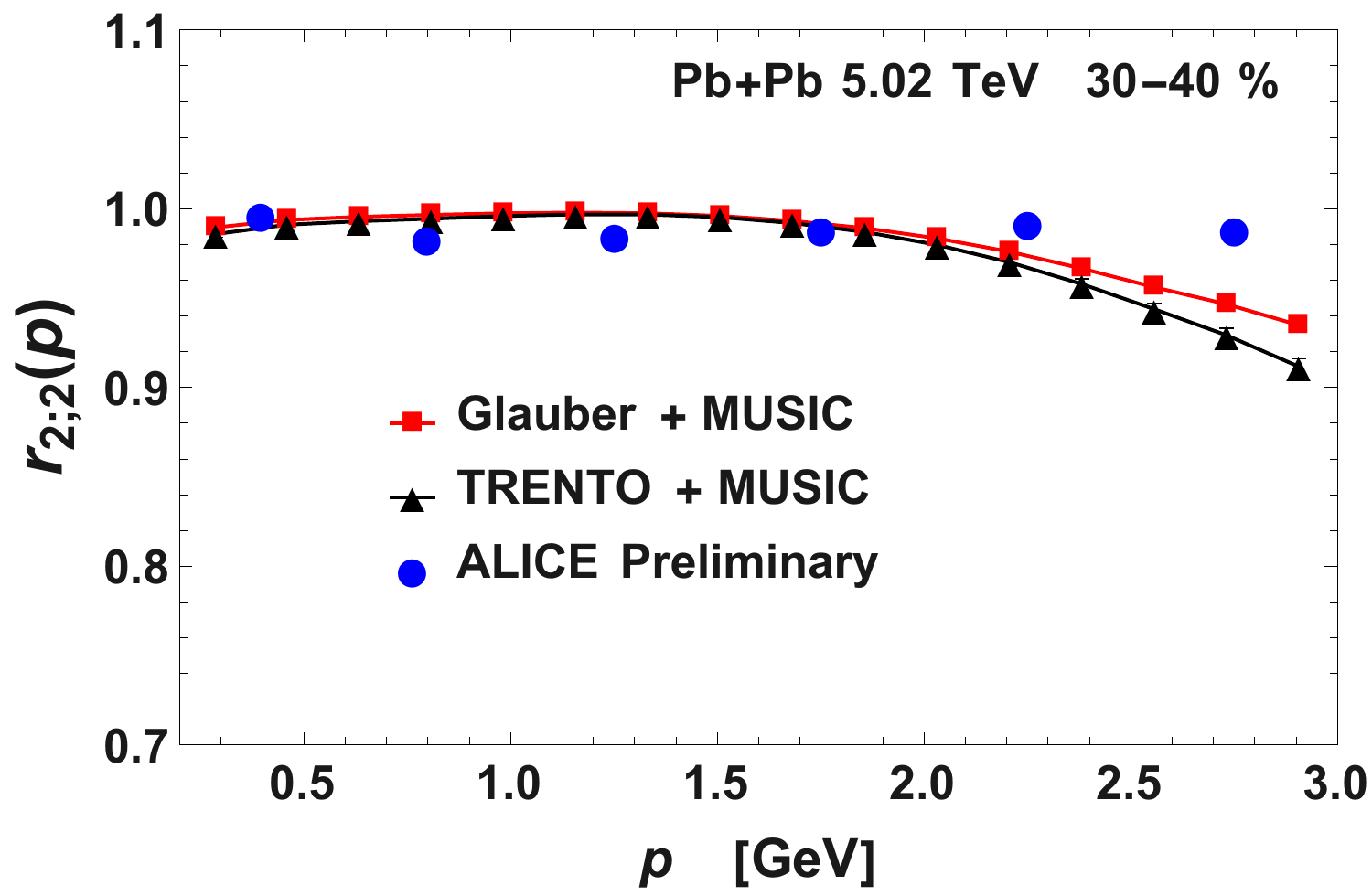} 
\end{center}
\caption{Same as Fig. \ref{fig:r2205} but for centrality $30-40$\%.}
\label{fig:r223040}
\end{figure}

The ALICE Collaboration has presented measurements of the harmonic
flow factorization breaking using only one bin in transverse momentum
\cite{ALICE:2017lyf,NielsenIS2021}.  That way the limitations from low multiplicity
in bins at high transverse momentum can be partly overcome.
The factorization breaking coefficient  is
defined as
\begin{equation}
r_n(p)=\frac{\langle V_n V_n^\star(p) \rangle}{\sqrt{\langle v_n^2 \rangle \langle v_n^2(p) \rangle}} 
\end{equation}
and it measures the correlation  of the harmonic flow vectors between
the momentum averaged  flow and the flow at the transverse momentum $p$.
The factorization breaking coefficient is also written as
\begin{equation}
r_n(p)=\frac{v_n\{2\}(p)}{v_n[2](p)} 
\end{equation}
and is a measure of the difference between two definitions \cite{Heinz:2013bua}
of the differential harmonic flow coefficient
  $v_n\{2\}(p)$ (Eq. \ref{eq:v22}) and 
 \begin{equation}
 v_n[2](p)=\sqrt{\langle V_n(p) V_n^\star(p)  \rangle} 
  \end{equation}

\begin{figure}
\vspace{5mm}
\begin{center}
\includegraphics[width=0.48 \textwidth]{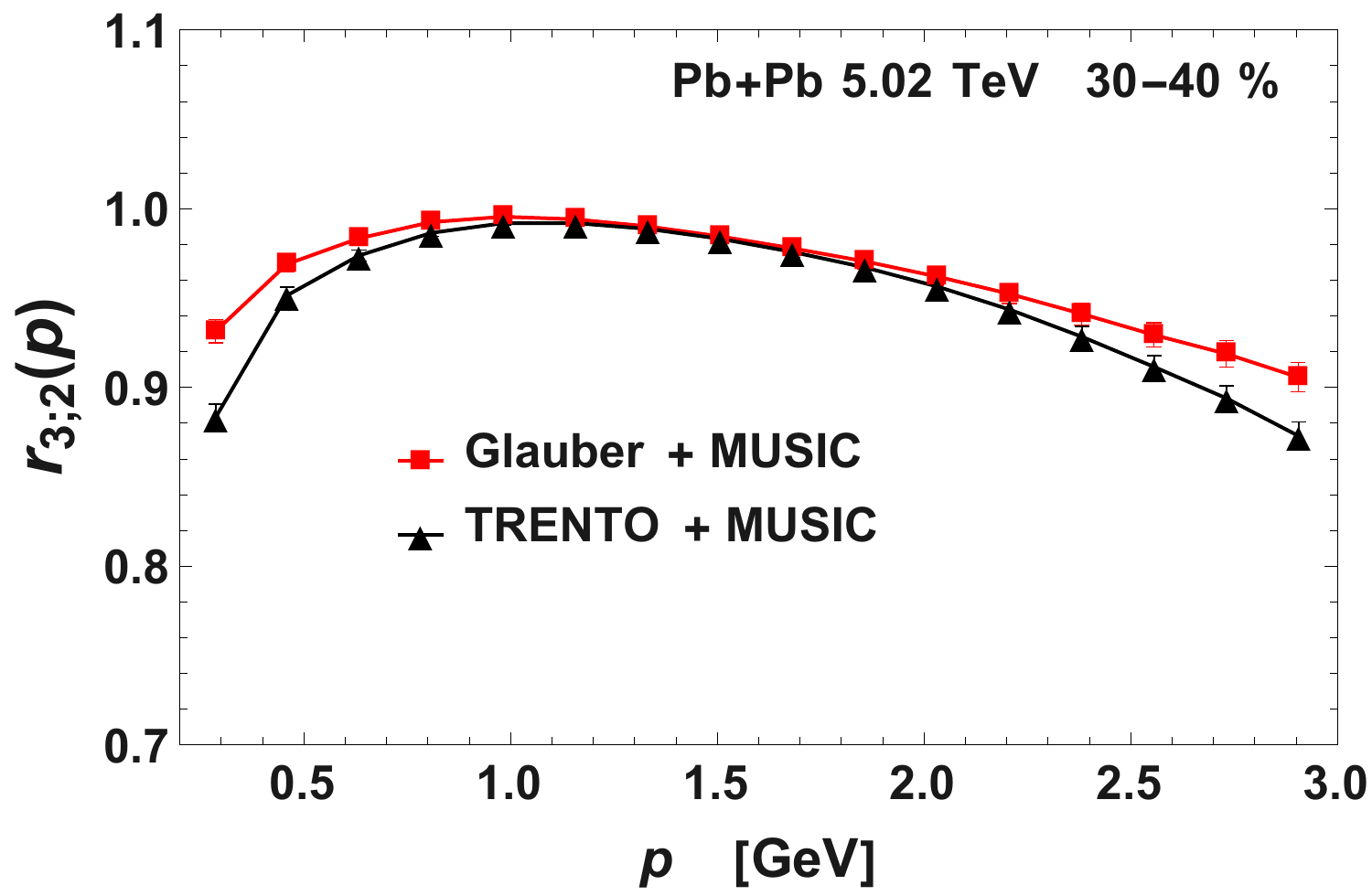} 
\end{center}
\caption{Same as Fig. \ref{fig:r223040} but for the triangular flow.}
\label{fig:r323040}
\end{figure}

The factorization breaking coefficient for the harmonic flow vector squared is
\begin{equation}
r_{n;2}(p)=\frac{\langle V_n^2 V_n^{\star }(p)^{2}\rangle}{\sqrt{\langle v_n^4 \rangle \langle v_n^4(p) \rangle}} \ .
\label{eq:r22}
\end{equation}
Note that the factorization breaking coefficient (\ref{eq:r22}) is
defined as the correlation coefficient between $V_n^2$ and $V_n(p)^2$.
In Figs. \ref{fig:r2205}, \ref{fig:r223040}, and \ref{fig:r323040} the results obtained in the hydrodynamic model are shown. The decorrelation of the flow vectors is the largest for centralities where fluctuations dominate,
i.e. for  the elliptic flow in central collisions and for the triangular flow at any centrality  (the results for the triangular flow at other centralities are qualitatively similar as for the centrality $30-40$\%, shown in Fig. \ref{fig:r323040}). The model results are similar to the data for central collisions (Fig. \ref{fig:r2205}). For semi-central events we find a much stronger decorrelation in the model for the flow vectors squared than in the data. The main difference shows up for $p>2.0$ GeV. The model calculations do not include any contribution from non-flow correlations, while the  experimental  data have non-flow effects reduced using rapidity gaps. We cannot  discriminate quantitatively in this study how much of the difference is due to remaining non-flow effects and how much can be attributed to genuine difference in flow  fluctuation between the model
and  experiment.

\begin{figure}
\vspace{5mm}
\begin{center}
\includegraphics[width=0.48 \textwidth]{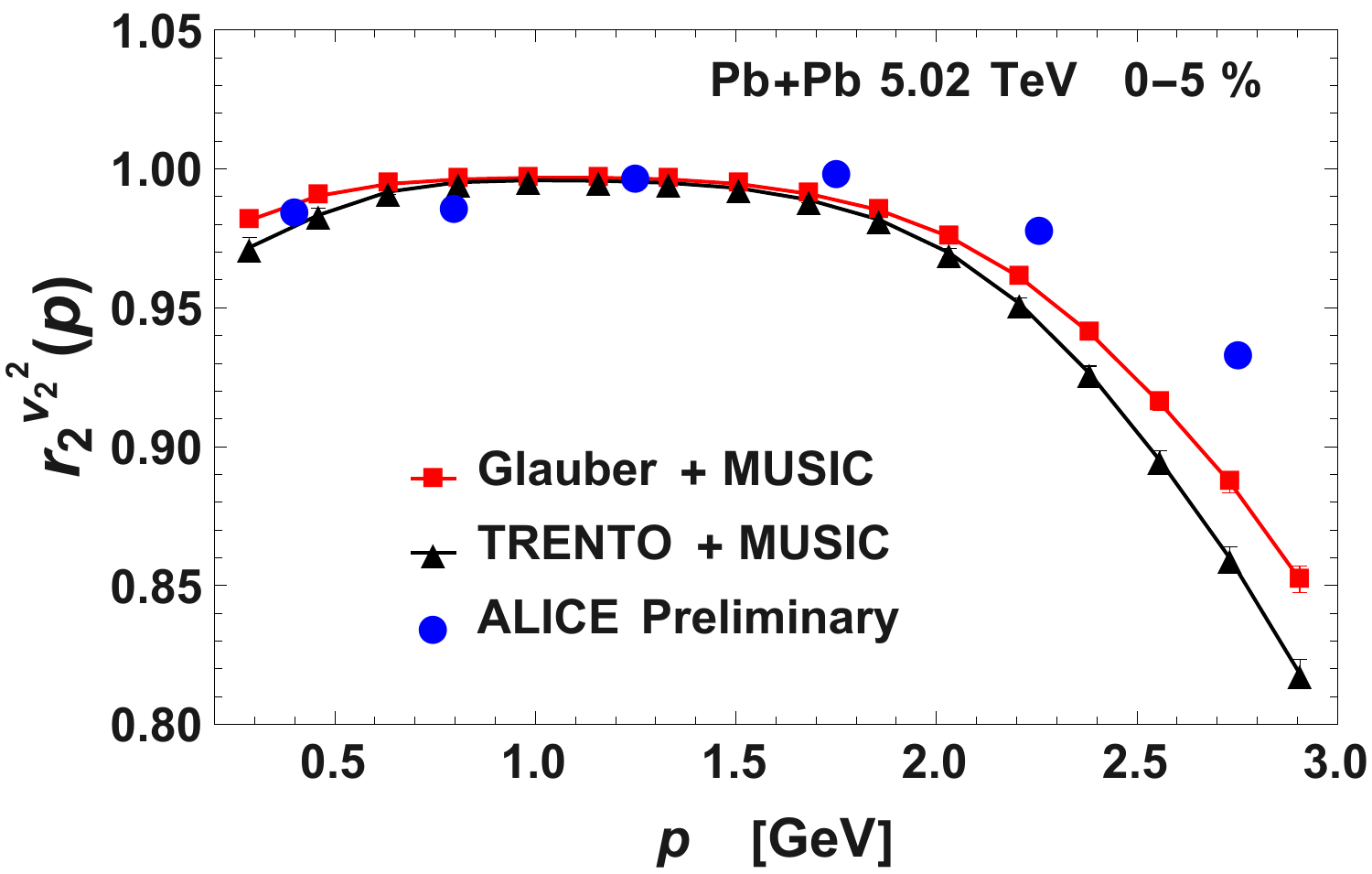} 
\end{center}
\caption{The factorization breaking coefficient between  the magnitudes of the elliptic flow squared $v_2(p)^2$ and $v_2^2$  as a
function of the transverse momentum $p$,
for Pb+Pb collisions with centrality $0-5$\%. The red squares and black triangles denote the results obtained with the Glauber model and the TRENTO model initial conditions respectively. The experimental data of the ALICE Collaboration are represented using the blue dots \cite{NielsenIS2021}.}
\label{fig:r22v05}
\end{figure}

\begin{figure}
\vspace{5mm}
\begin{center}
\includegraphics[width=0.48 \textwidth]{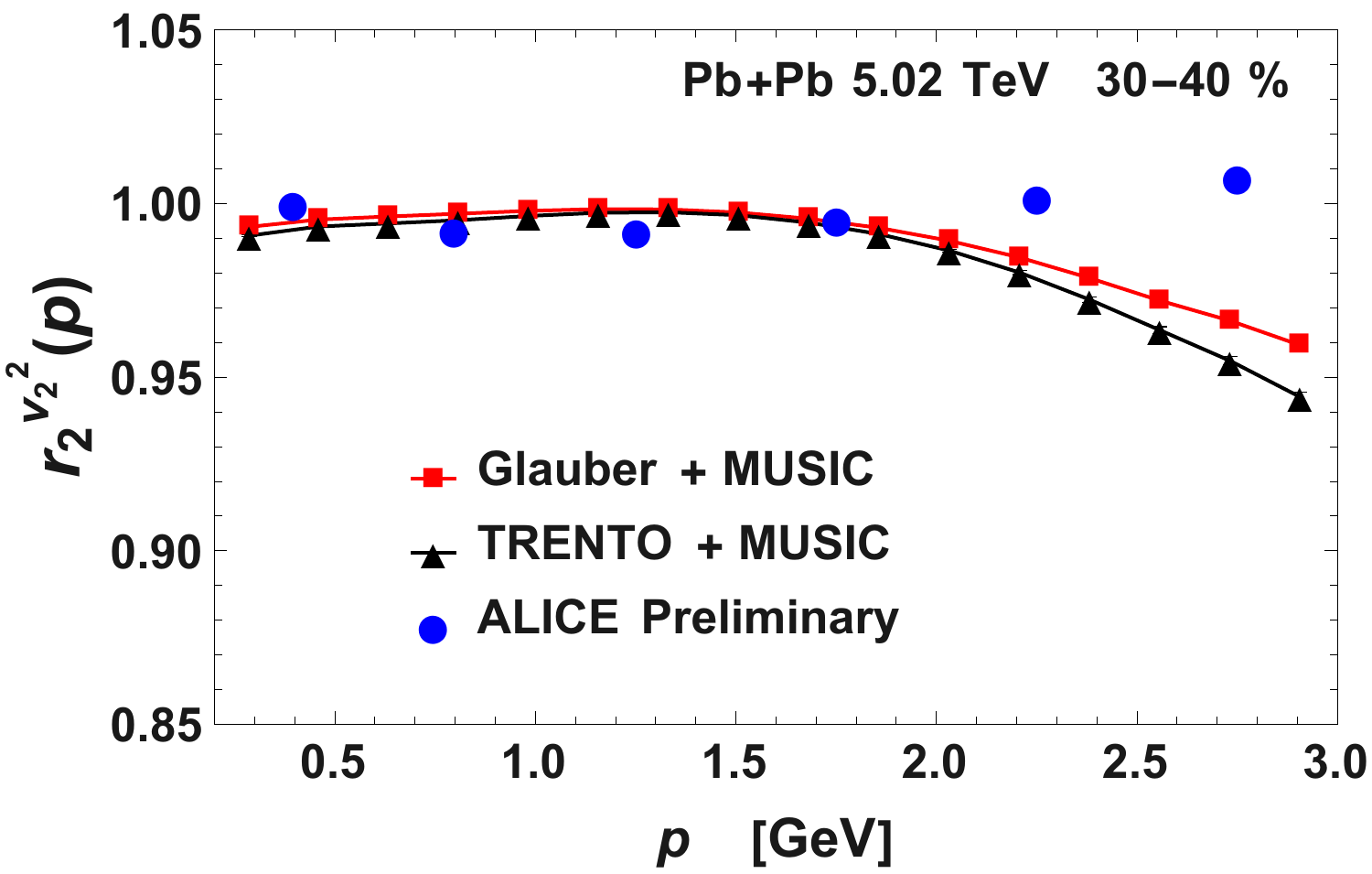} 
\end{center}
\caption{Same as Fig. \ref{fig:r22v05} but for centrality $30-40$\%.}
\label{fig:r22v3040}
\end{figure}


With a  four-particle correlator one can  measure the factorization breaking coefficient for
the square of the magnitude of the harmonic flow~:
\begin{equation}
r_{n}^{v_n^2}(p)=\frac{\langle v_n^2 v_n^{ 2}(p)\rangle}{\sqrt{\langle v_n^4 \rangle \langle v_n^4(p) \rangle}} \ .
\label{eq:mag2}
\end{equation}
In Figs. \ref{fig:r22v05} and  \ref{fig:r22v3040}, are shown the
results obtained for the elliptic flow in the hydrodynamic model. The results for the triangular flow are qualitatively similar (not shown).
We observe that the magnitude decorrelation accounts for roughly one half of the flow vector decorrelation
\begin{equation}
 \left[1- r_{n;2}(p)\right] \simeq 2 \left[\ 1- r _{n}^{v_n^2}(p) \ \right] \ .
\label{eq:r22v}
\end{equation}
This means that the remaining angle decorrelation accounts roughly for the other half of the total vector decorrelation. While in general, any proportion between angle and magnitude decorrelation is possible, we note that previous experimental measurements  \cite{ATLAS:2017rij} and model calculations \cite{Bozek:2017qir} have observed an approximately equal strength of angle and magnitude decorrelation for flow correlations in rapidity.
Similar to the flow vector factorization breaking coefficients, also for the  magnitude factorization breaking coefficient, the data lie significantly above the simulations at high transverse momentum, which may be partly due to the  contribution of
non-flow correlations. This discrepancy is particularly pronounced for the $30-40$\% centrality. In fact, some of the data points lie above one, suggesting the dominance of non-flow correlations.

A comment is in order on  how the  flow vectors squared
and flow magnitudes squared factorization breaking coefficients
are extracted from the experimental data of the ALICE Collaboration.
The experiment does not   measure directly the flow vectors and the flow magnitudes
factorization coefficients  (Eqs. \ref{eq:r22} and \ref{eq:mag2}) \cite{NielsenIS2021}. The difficulty lies in the extraction of the four-particle correlator in the denominator of the correlation formulae
$ \langle  v_n^4(p)\rangle$, 
with all four particles in a narrow transverse momentum bin. On the other hand the fourth moment $\langle v_n^4 \rangle$ can be measured.  The ALICE
Collaboration presents the results for the correlators with different scaling, e.g. for flow magnitude squared correlations,
\be
\frac{\langle v_n^2 v_n^{ 2}(p)\rangle}{{\langle v_n^2 \rangle \langle v_n^2(p) \rangle}} \ .
\ee
By dividing this scaled correlator by $\langle v_n^4\rangle / \langle v_n^2\rangle^2$ \cite{NielsenIS2021} an estimate of the flow vector squared  correlation
\be
r_{n;2}(p) \simeq \frac{\langle V_n^2 V_n^{\star  }(p)^{2}\rangle \langle v_n^2 \rangle }{{\langle v_n^4 \rangle \langle v_n^2(p) \rangle}} 
\label{eq:r22app}
\ee
or the flow magnitude squared correlation
\be
r_{n}^{v_n^2}(p) \simeq \frac{\langle v_n^2 v_n^{ 2}(p)\rangle \langle v_n^2 \rangle }{{\langle v_n^4 \rangle \langle v_n^2(p) \rangle}} 
\label{eq:mag2app}
\ee
can  be obtained. The difference between Eqs. (\ref{eq:r22app}) and
(\ref{eq:mag2app})  and the factorization breaking   coefficients defined in Eqs.
(\ref{eq:r22}) and (\ref{eq:mag2}) is a factor
\be
\sqrt{\frac{\langle v_n^4(p) \rangle \langle v_n^2\rangle^2  }{\langle v_n^4 \rangle \langle v_n^2(p)\rangle^2 }} \ .
\ee
  We have checked that in the hydrodynamic model the deviation of this factor from $1$ is less than $6 \times 10^{-3}$ for $0.5$~GeV $< p < 3.0$~GeV.
  The experimental data for the flow vector and magnitude correlation shown in the Figures
  are obtained using the formulae (\ref{eq:r22app}) and (\ref{eq:mag2app}).

\section{Flow angle decorrelation}

\begin{figure}
\vspace{5mm}
\begin{center}
\includegraphics[width=0.48 \textwidth]{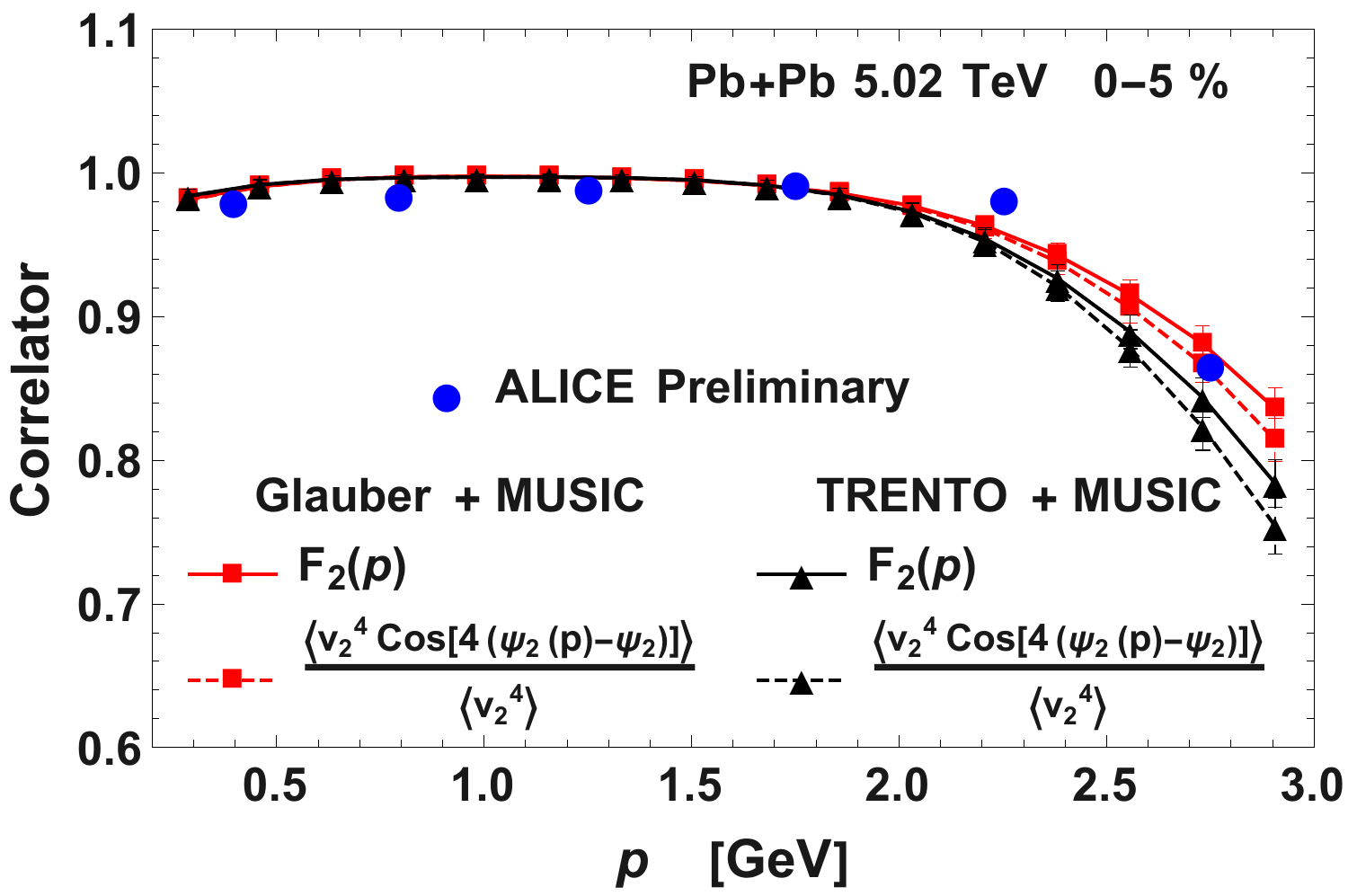} 
\end{center}
\caption{The flow angle correlation estimated as the ratio of the flow vector and the flow magnitude factorization breaking coefficients  (Eq. \ref{eq:ang}) for the elliptic flow, as a
function of the transverse momentum $p$,
for Pb+Pb collisions with centrality $0-5$\%. The red squares and black triangles denote the results obtained with the Glauber model and the TRENTO model initial conditions respectively. The experimental data of the ALICE Collaboration are represented using  blue dots \cite{NielsenIS2021}. The symbols with the dashed lines denote the angle correlations calculated directly from the simulated events (Eq. \ref{eq:angmodel}).}
\label{fig:ang2205}
\end{figure}

\begin{figure}
\vspace{5mm}
\begin{center}
\includegraphics[width=0.48 \textwidth]{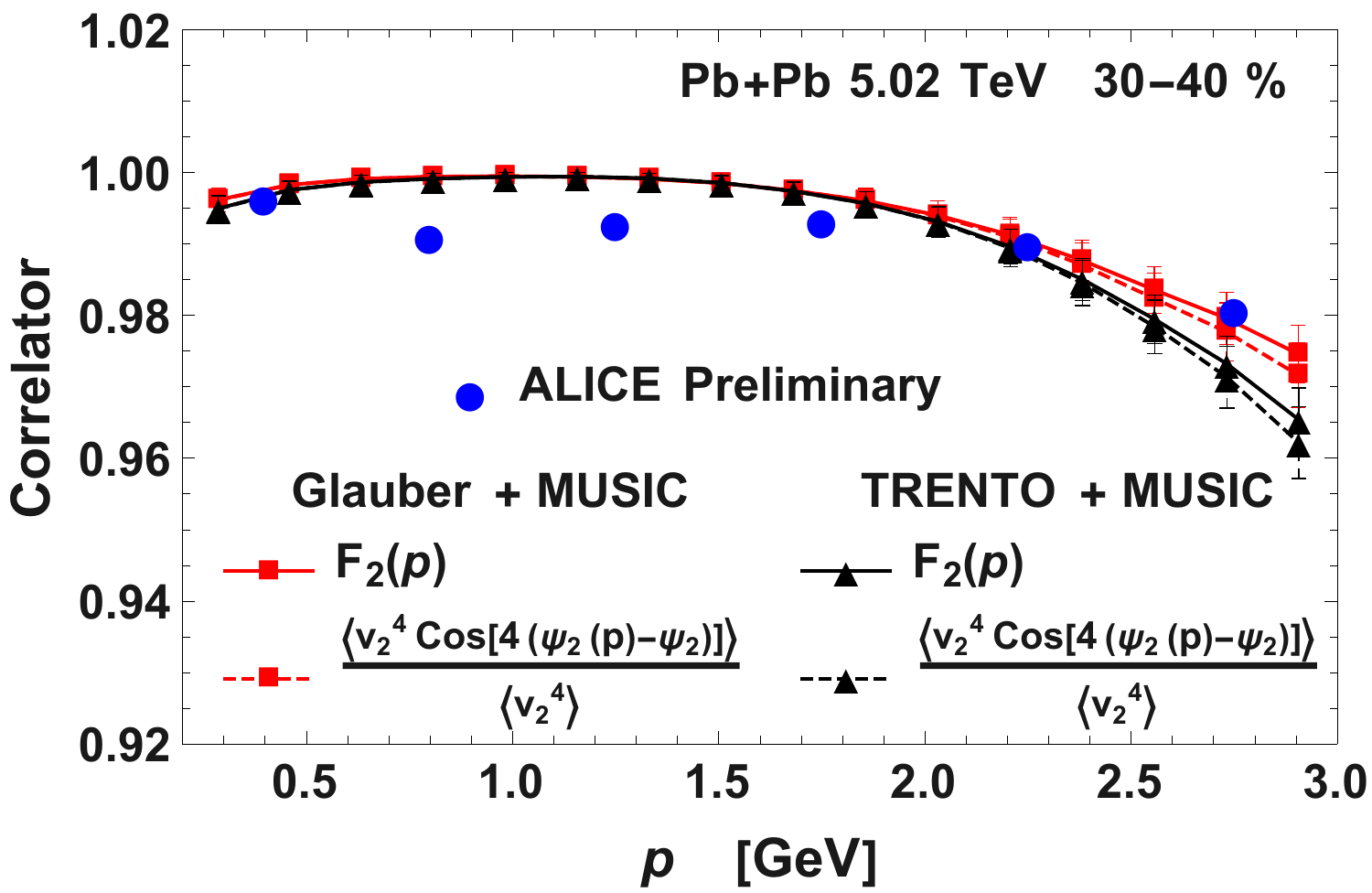} 
\end{center}
\caption{Same as Fig. \ref{fig:ang2205} but for centrality $30-40$\%.}
\label{fig:ang223040}
\end{figure}

\begin{figure}
\vspace{5mm}
\begin{center}
\includegraphics[width=0.48 \textwidth]{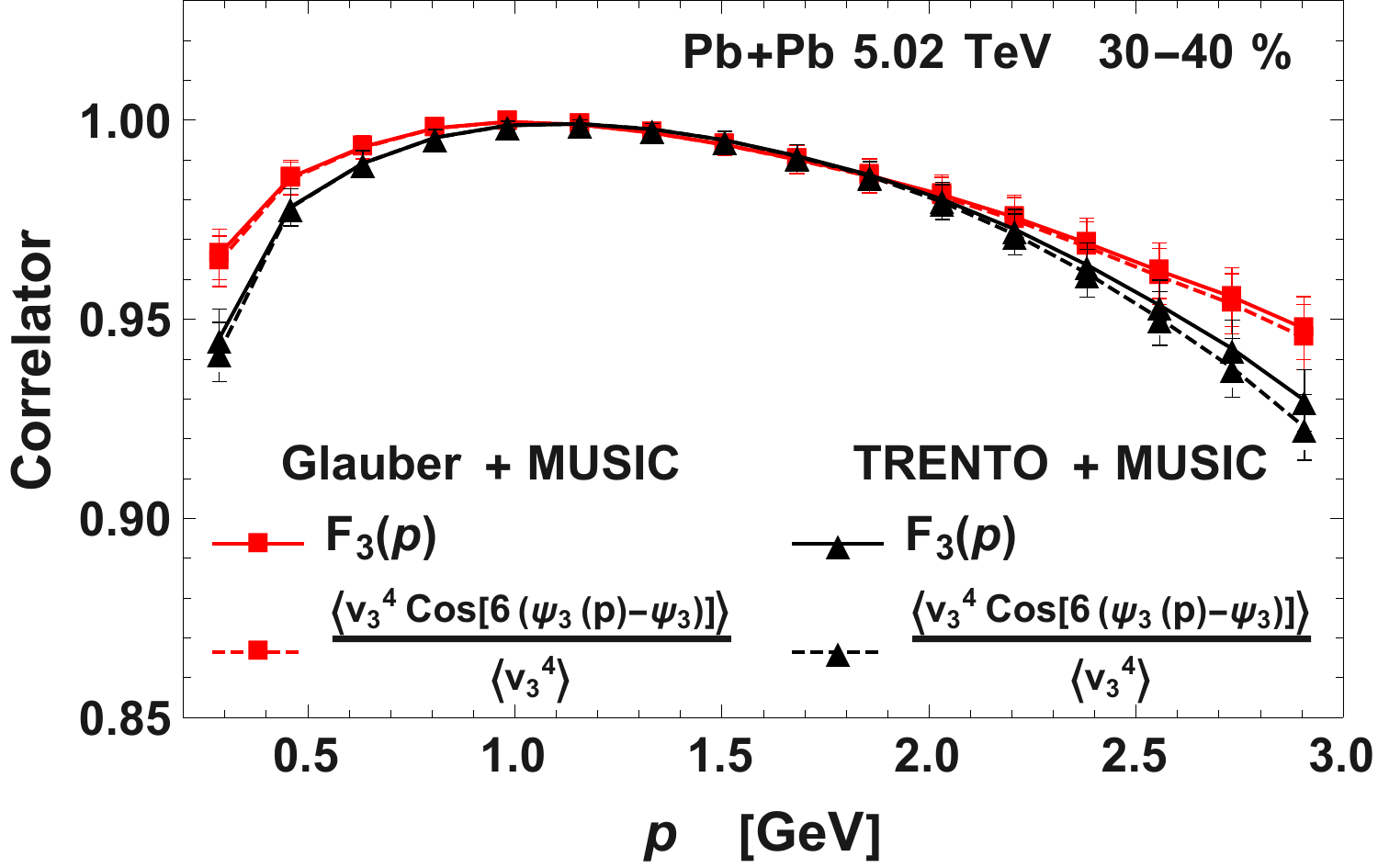} 
\end{center}
\caption{Same as Fig. \ref{fig:ang223040} but for the triangular flow.}
\label{fig:ang333040}
\end{figure}

An experimental observable directly measuring the flow angle correlation cannot be defined. 
An estimate of the flow angle correlation can be obtained as the ratio of a four-particle
flow vector correlator and a flow magnitude correlator \cite{Jia:2017kdq,NielsenIS2021}. The correlator measure used by the ALICE Collaboration for the correlation of the flow angle in transverse momentum is
\begin{equation}
F_{n}(p)=\frac{\langle V_n^2 V_n^{\star }(p)^{2}\rangle}
{\langle v_n^2 v_n^{ 2}(p)\rangle} \ .
\label{eq:ang}
\end{equation}
It is  the ratio of the flow vector squared (\ref{eq:r22}) and flow
magnitude squared (\ref{eq:mag2}) factorization breaking coefficients.
In Figs. \ref{fig:ang2205}, \ref{fig:ang223040}, and \ref{fig:ang333040}
are shown the results of the hydrodynamic simulations compared to the  ALICE Collaboration data.
The model simulations predict  a noticeable  flow angle decorrelation between the global flow $V_n$ and the differential, momentum dependent flow $V_n(p)$, both for the elliptic and triangular flows.
The preliminary data of ALICE Collaboration are consistent with the simulation for the elliptic flow in central collisions. For the centrality $30-40$\% (Fig. \ref{fig:ang223040}),
the simulation and the experiment show a  smaller angle decorrelation than in central collisions. Non-flow correlations are expected to be relatively more important in that case.  The difference in the strength of the decorrelation  between central and semi-central collisions  comes from the global correlation of the elliptic flow due to the initial geometry in non-central collisions.

The  formula in  Eq. (\ref{eq:ang}) is a way to estimate the
flow angle decorrelation in experiment. It measures
\begin{equation}
\frac{\langle  v_n^2 v_n(p)^2 \cos\left[ 2 n  \left( \Psi_n(p) -\Psi_n \right)\right] \rangle}{\langle v_n^2 v_n(p)^2 \rangle} \ .
\label{eq:angexp}
\end{equation}
Its use as an estimate of the flow angle decorrelation
\begin{equation}
\frac{\langle  v_n^4 \cos\left[ 2 n  \left( \Psi_n(p) -\Psi_n \right)\right] \rangle}{\langle v_n^4 \rangle} \ ,
\label{eq:angmodel}
\end{equation}
 involving only the angle decorrelation
is based on the implicit assumption
that the  decorrelations of the flow magnitude in the numerator and
denominator cancel. 
The above  formula (\ref{eq:angmodel}) cannot be directly applied for the experimental analysis, but in the model we can check whether the two formulae (\ref{eq:angexp}) and (\ref{eq:angmodel})
give similar results. In other worlds, we check if the momentum dependence of the magnitude and flow angle in the numerator of Eq. (\ref{eq:angexp}) factorizes in the hydrodynamic model simulations. The results presented
in Figs. \ref{fig:ang2205}, \ref{fig:ang223040}, and \ref{fig:ang333040}
indicate
that the two formulae give similar results, suggesting that  the experimental measure
(\ref{eq:ang}) could  be used to estimate the weighted flow angle decorrelation.
 Please note that the angle correlation is weighted
by the fourth power of  flow magnitude. Only the $v_n^4$ weighted flow angle decorrelation is measured. We have checked in our simulation, that  using a simple average
\begin{equation}
\langle  \cos\left[ 2 n  \left( \Psi_n(p) -\Psi_n \right)\right] \rangle
\label{eq:angsimple}
\end{equation}
instead, would give a very different result. This results are not unexpected. In events with large flow (large $|V_n|$ and large $|V_n(p)|$) the random decorrelation between the two vector is relatively smaller, and the angle decorrelation is small. For a detailed discussion of the effect we refer the reader to Ref. 
\cite{Bozek:2017qir}. Please note, that the simple angle correlation (\ref{eq:angsimple}) cannot be measured experimentally.

\begin{figure}
\vspace{5mm}
\begin{center}
\includegraphics[width=0.48 \textwidth]{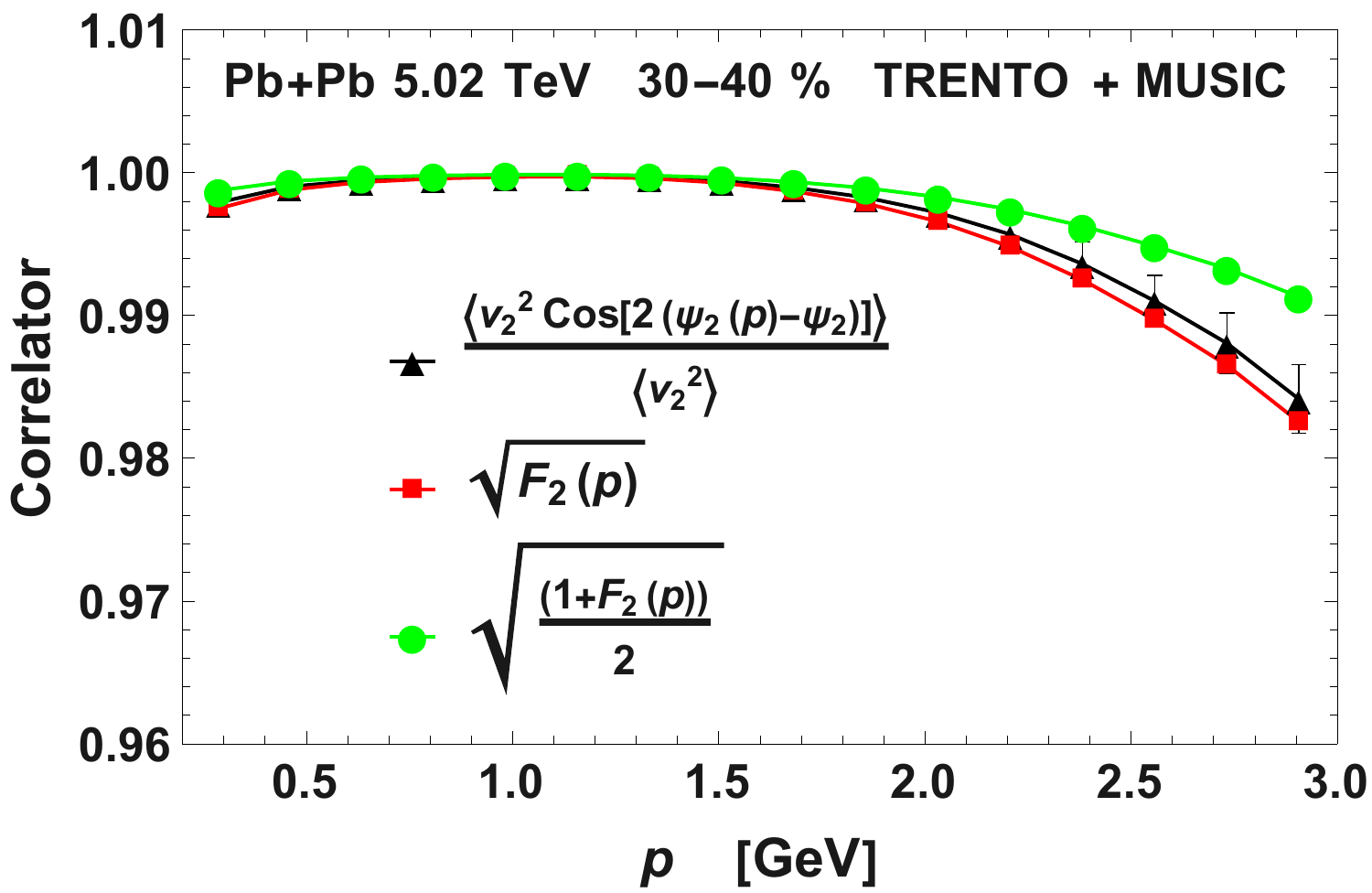} 
\end{center}
\caption{The flow angle correlation between flow vectors $V_2(p)$ and $V_2$    as a
function of the transverse momentum $p$, for Pb+Pb collisions with centrality $30-40$\% (black triangles), compared to its approximation by $\sqrt{F_2(p)}$ (red squares) and its upper limit $\sqrt{(1+F_2(p))/2}$ (green dots).}
\label{fig:angcomp}
\end{figure}

In the experiment, only the angle decorrelation between
the flow vectors {\it squared } can be measured. In the model one can check how it is related to the angle correlation between  first moments of flow vectors 
\be
\frac{\langle v_n^2 \cos\left[ n \left( \Psi_n(p)-\Psi_n\right) \right]\rangle}{\langle v_n^2\rangle} \ .
\ee
We find that the flow angle correlation between the first moment can be approximated (Fig. \ref{fig:angcomp}) as the squared root of the angle decorrelation for the flow vectors squared
\be
\frac{\langle v_n^2 \cos\left[ n \left( \Psi_n(p)-\Psi_n\right)\right]\rangle}{\langle v_n^2\rangle} \simeq \sqrt{F_n(p)} \ .
\ee
A similar relation was found for  correlators
of higher moment of flow vectors
in pseudorapidity \cite{ATLAS:2017rij}. In Fig. \ref{fig:angcomp} the upper limit for the flow angle correlation $\sqrt{(1+F_n(p))/2}$, proposed by the ALICE Collaboration \cite{NielsenIS2021}, is also shown.

\section{Mixed harmonics}

Correlations between event planes for mixed  flow harmonics 
are interesting as a measure of nonlinearities in the hydrodynamic
expansion
or of correlations in the initial
state \cite{Bhalerao:2011yg,Teaney:2012ke,Luzum:2013yya,Jia:2012ma,Jia:2012ju,Teaney:2013dta,Qiu:2012uy,Jia:2012sa,Bhalerao:2013ina,Qian:2016fpi,Giacalone:2016afq,ALICE:2017fcd}.  Quantities discussed in these studies are based on correlators  of harmonic flow vectors of different orders.
Typically these moments involve the momentum averaged harmonic flow.
The momentum dependence of the nonlinear response \cite{Qian:2017ier} or the covariance \cite{Bozek:2017thv}  between
flow  harmonics of different orders  have been discussed in models,
using moments of mixed flow harmonics 
in bins of transverse momentum. Such studies  of many-particle
correlators in small bins 
cannot be easily performed experimentally due to limited statistics.

\begin{figure}
\vspace{5mm}
\begin{center}
\includegraphics[width=0.48 \textwidth]{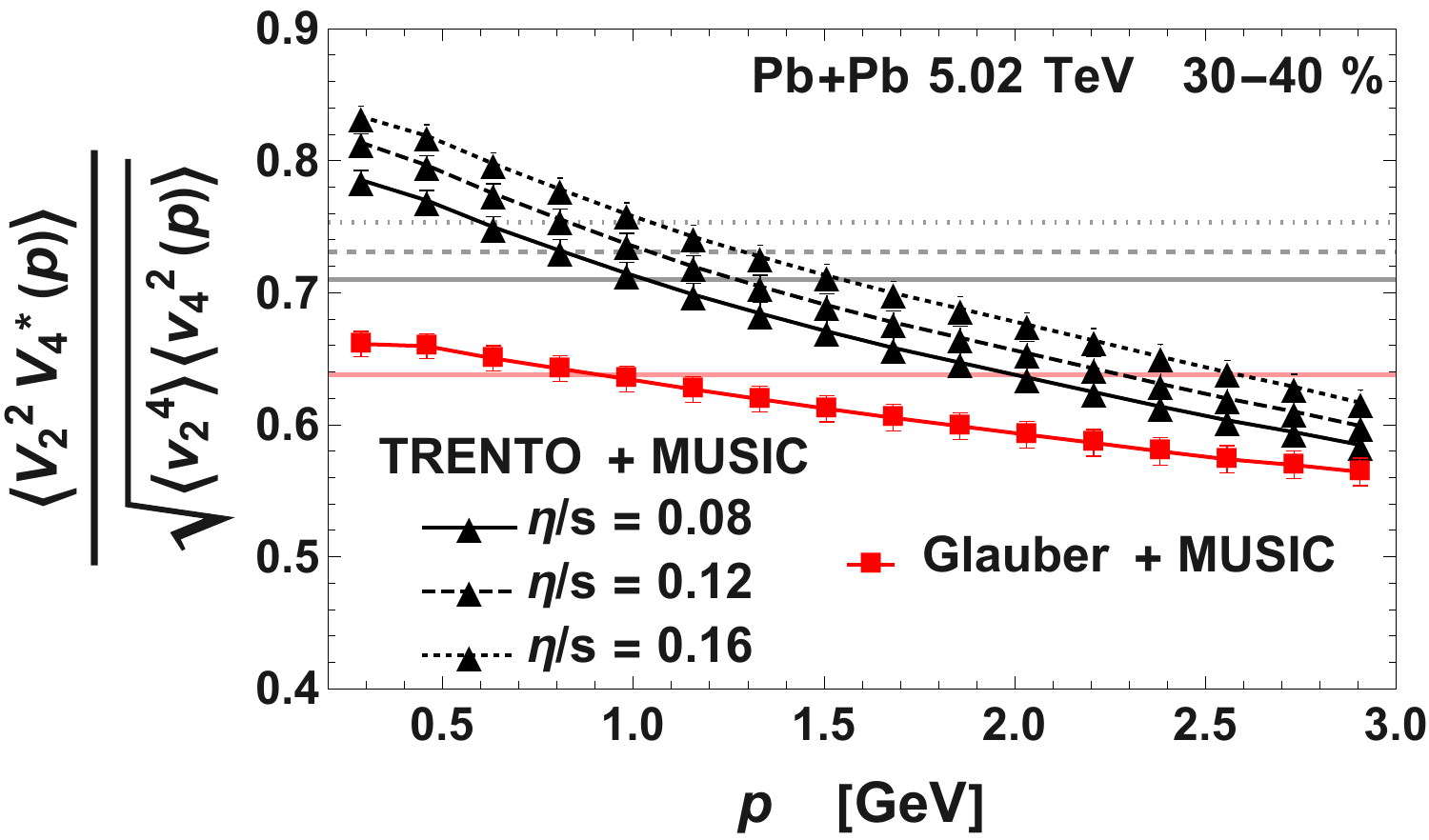} 
\end{center}
\caption{The correlation coefficient between $V_2^2$ and $V_4(p)$ as  a function of the transverse momentum $p$ in Pb+Pb collisions with centrality $30-40$\%.
The results obtained in the hydrodynamic model using Glauber model and TRENTO model initial conditions are denoted with red squares and black triangles respectively. For the TRENTO model initial conditions,   the solid, dashed  and dotted lines represent results with $\eta/s=0.08,\ 0.12, \ 0.16$ respectively.
The horizontal lines represent the corresponding correlation coefficients  for momentum averaged flow vector $V_2^2$ and $V_4$.}
\label{fig:v22v43040}
\end{figure}

\begin{figure}
\vspace{5mm}
\begin{center}
\includegraphics[width=0.48 \textwidth]{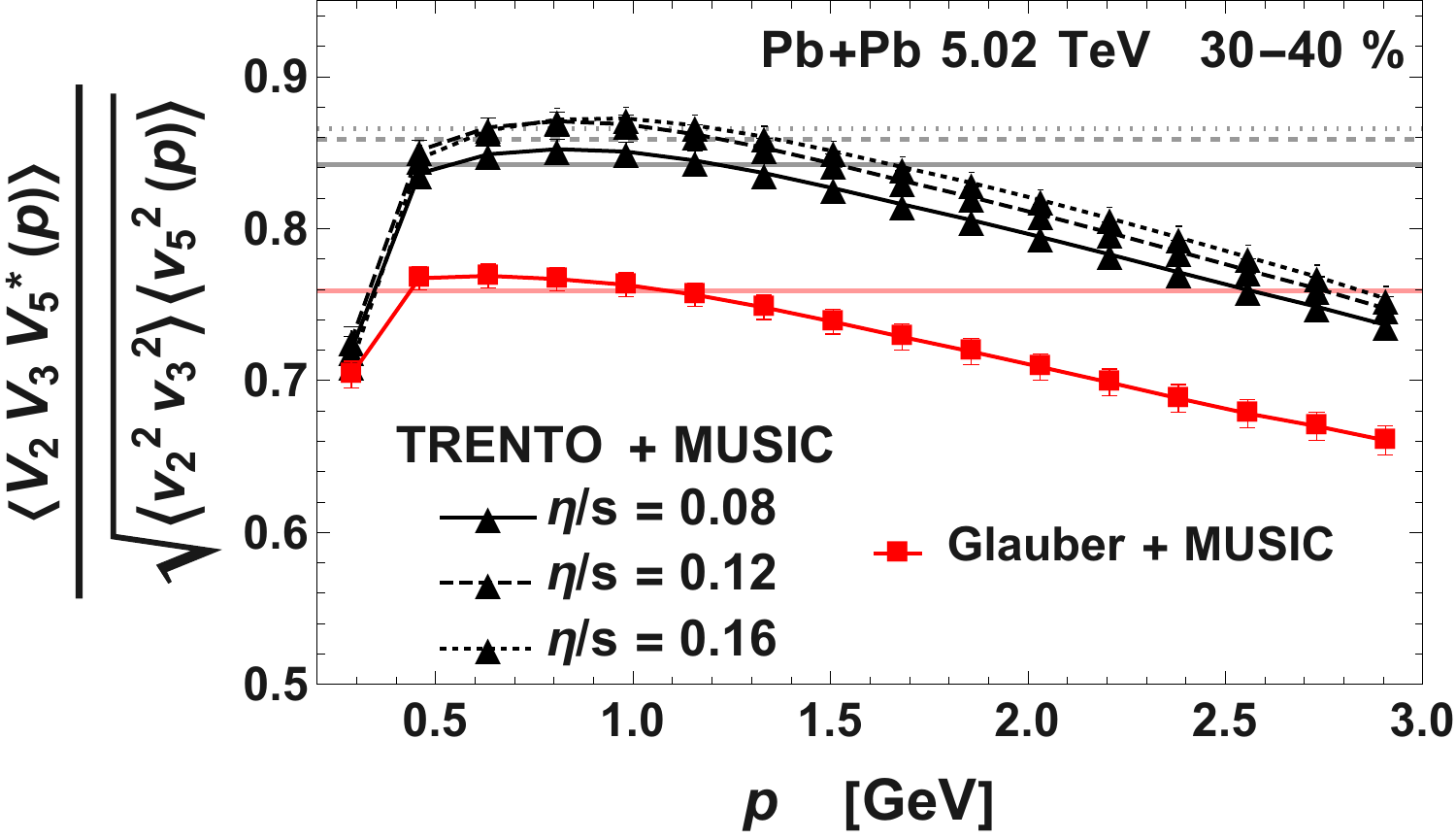} 
\end{center}
\caption{Same as in Fig. \ref{fig:v22v43040} but for the correlation
coefficient between $V_5(p)$ and $V_3V_2$.}
\label{fig:v5323040}
\end{figure}

\begin{figure}
\vspace{5mm}
\begin{center}
\includegraphics[width=0.48 \textwidth]{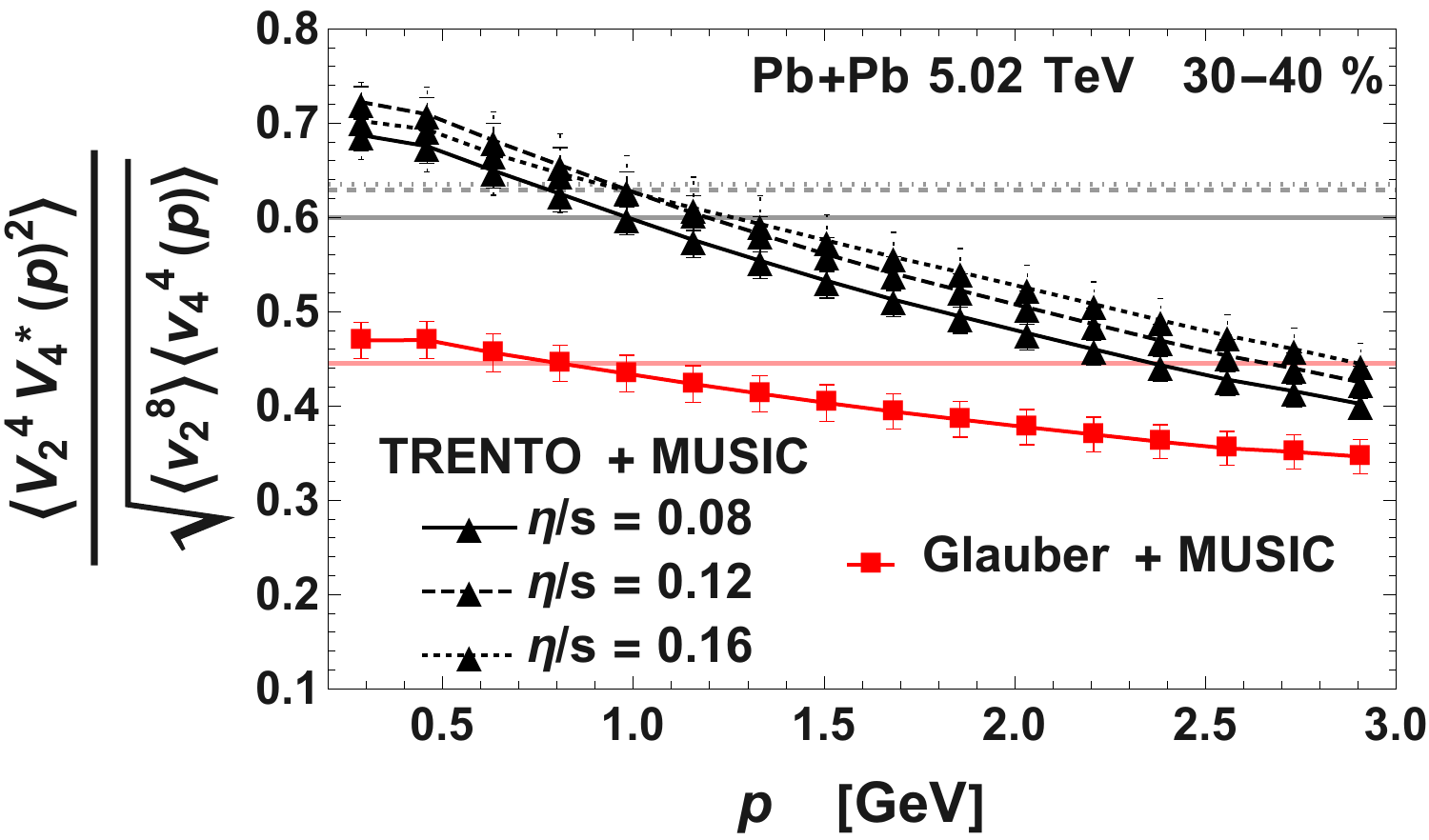} 
\end{center}
\caption{The correlation coefficient between the powers of flow harmonics $V_2^4$ and $V_4(p)^2$ as  a function of the transverse momentum $p$, for Pb+Pb collisions with centrality $30-40$\%.
The results obtained in the hydrodynamic model using Glauber model and TRENTO model initial conditions are denoted with red squares and black triangles respectively. For the TRENTO model initial conditions,   the solid, dashed and dotted lines represent results with $\eta/s=0.08,\ 0.12, \ 0.16$ respectively.
The horizontal lines represent the corresponding correlation coefficients for momentum averaged flow vectors $V_2^4$ and $V_4^2$.}
\label{fig:v24v423040}
\end{figure}


\begin{figure}
\vspace{5mm}
\begin{center}
\includegraphics[width=0.48 \textwidth]{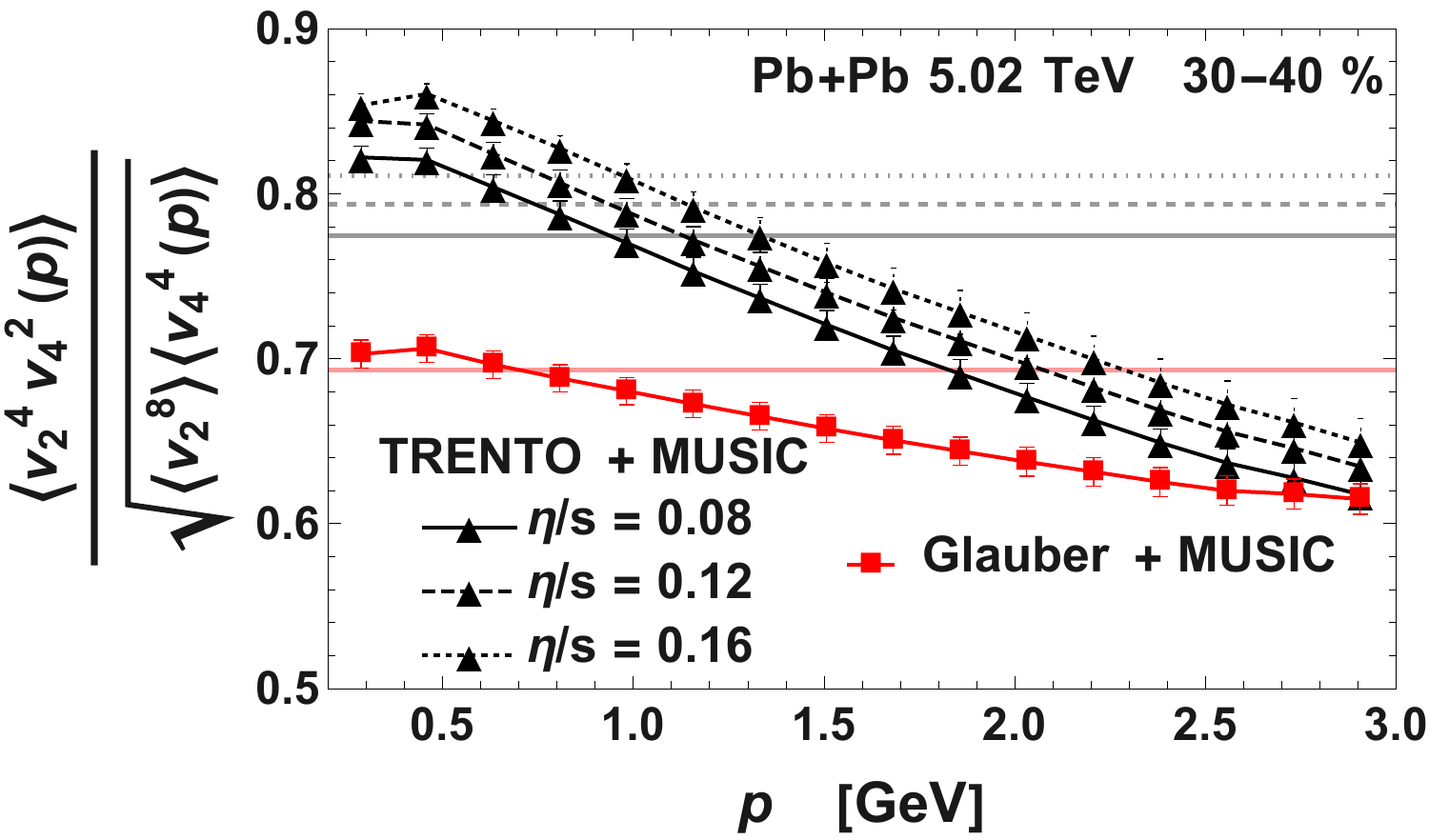} 
\end{center}
\caption{Same as in Fig. \ref{fig:v24v423040} but for the factorization breaking  coefficient between the powers of flow magnitudes $|V_4(p)|^2$ and $|V_2|^4$.}
\label{fig:mag24v423040}
\end{figure}


A correlation coefficient  (in this context the name correlation coefficient is used in the literature instead of the factorization breaking coefficient)
for mixed flow harmonics can be defined  with only one of the flow
harmonics restricted to a transverse momentum bin
\begin{equation}
\frac{\langle V_m^\star(p) V_k V_n \rangle}{\sqrt{\langle v_m(p)^2\rangle \langle v_k^2 v_n^2  \rangle}} \ ,
\end{equation}
with $m=k+n$.
In Fig. \ref{fig:v22v43040}, are shown the results for the correlation coefficient measuring the nonlinear coupling between $V_4(p)$ and $V_2^2$
as a  function   of the transverse momentum.  We present the correlation coefficient for  semi-central collisions only, where the nonlinear component in $V_4$ is dominant.
The dependence on transverse momentum gives an additional constraint on the initial state and on the hydrodynamic evolution. The correlation is the largest for small transverse momenta. In Fig. \ref{fig:v5323040}, is shown the analogous result for the coupling between $V_5$ and $V_3 V_2$. The results are qualitatively similar as for the $V_4$-$V_2^2$ coupling.

\begin{figure}
\vspace{5mm}
\begin{center}
\includegraphics[width=0.48 \textwidth]{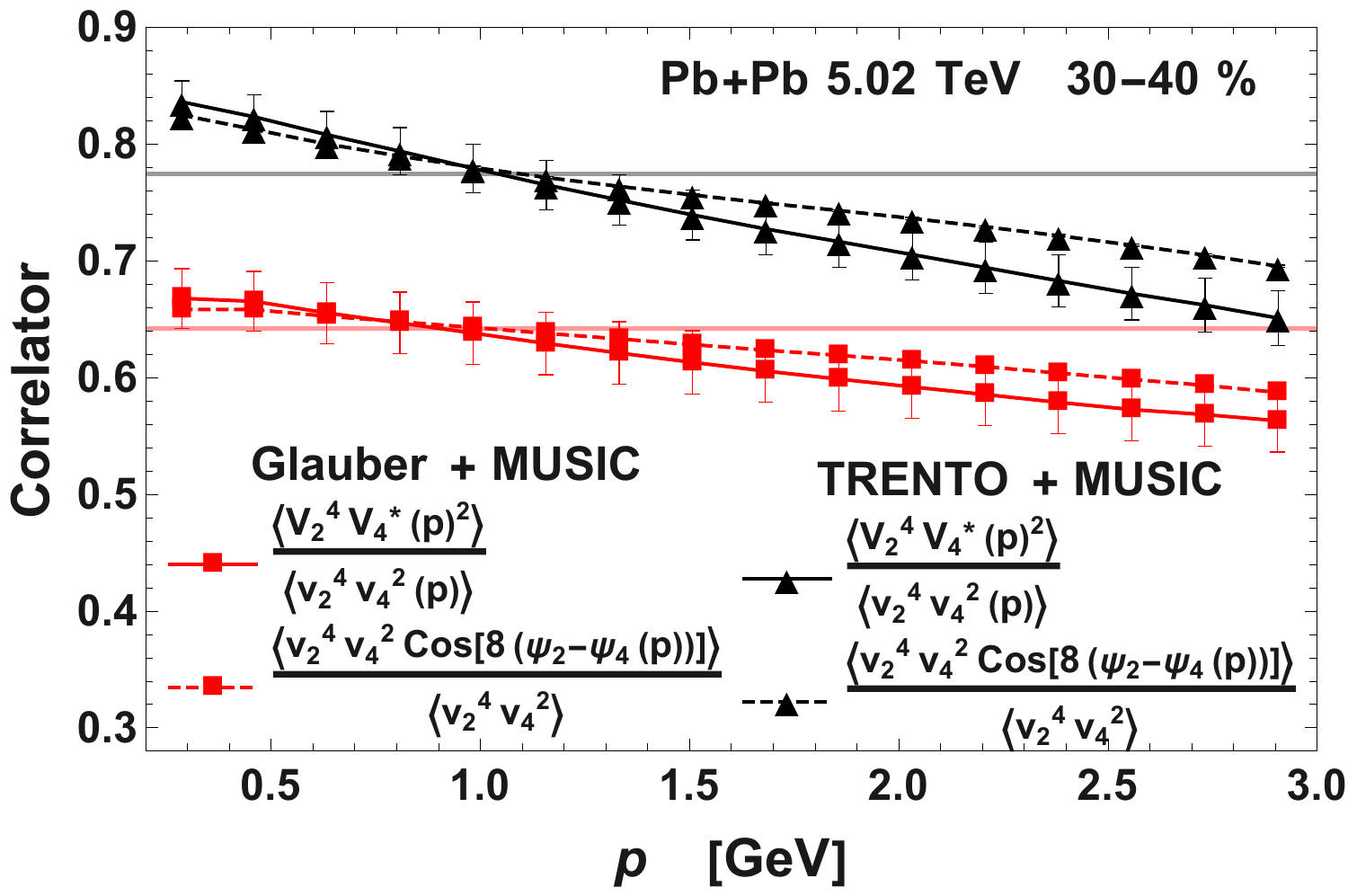} 
\end{center}
\caption{The flow angle decorrelation between $V_2^4$ and $V_4(p)^2$  as a
function of the transverse momentum $p$, for Pb+Pb collisions with centrality $30-40$\%.
The results obtained in the hydrodynamic model using Glauber model and TRENTO model initial conditions are denoted with red squares and black triangles respectively. The symbols with the dashed lines denote the angle correlations calculated directly from the simulated events (Eq. \ref{eq:angmodel422}).
The horizontal lines represent the corresponding angle decorrelation  for momentum averaged flow vectors $V_2^4$ and $V_4^2$.}
\label{fig:ang22v43040}
\end{figure}


\begin{figure}
\vspace{5mm}
\begin{center}
\includegraphics[width=0.48 \textwidth]{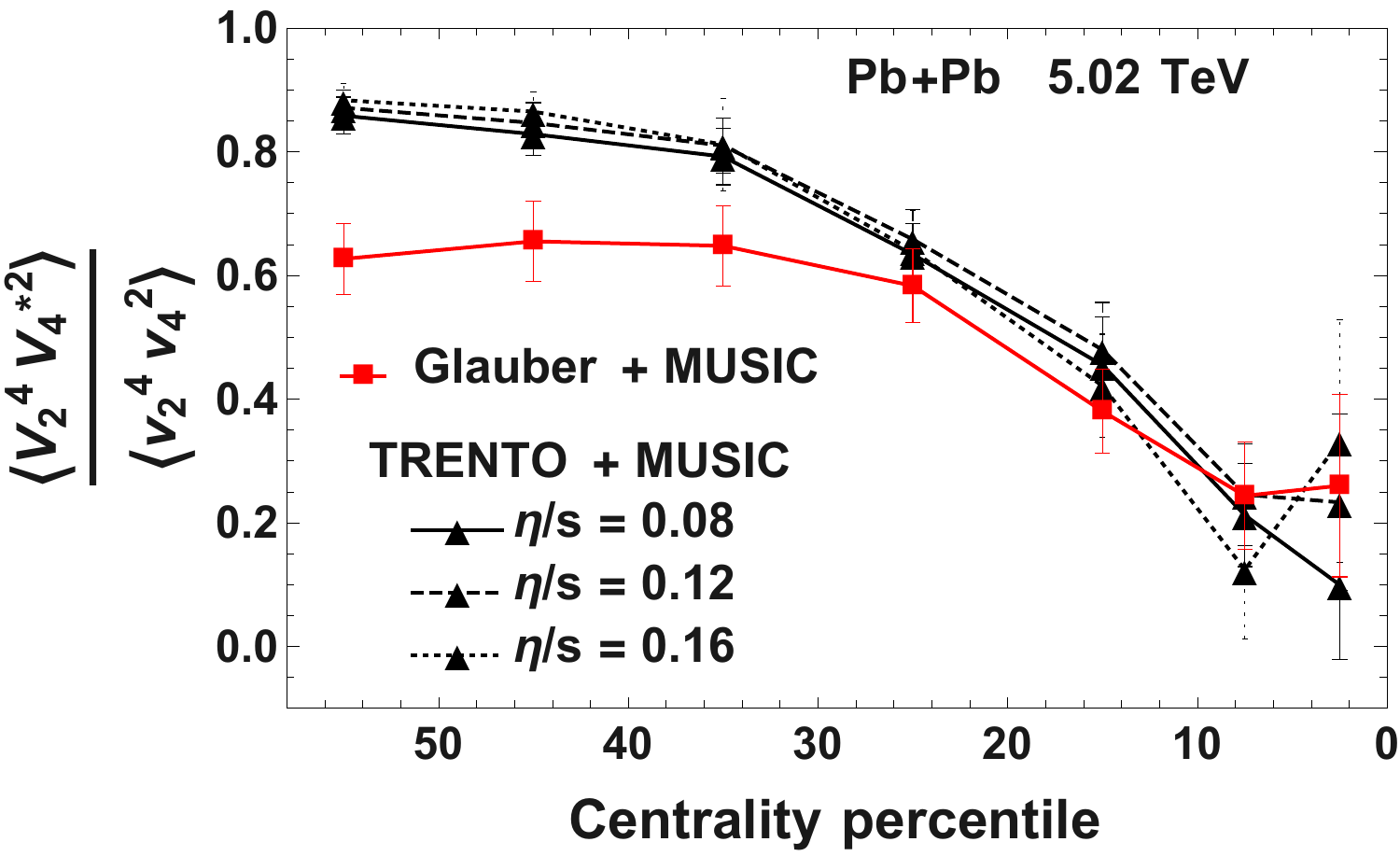} 
\end{center}
\caption{The flow angle correlation between $V_2^4$ and $V_4^2$  as 
a function of centrality in Pb+Pb collisions.
The results obtained in the hydrodynamic model using Glauber model and TRENTO model initial conditions are denoted with red squares and black triangles respectively. For the TRENTO model initial conditions,   the solid, dashed  and dotted lines represent results with $\eta/s=0.08,\ 0.12, \ 0.16$ respectively.}
\label{fig:angcentr}
\end{figure}

In order to extract the flow angle correlation  for mixed flow harmonic separately from the magnitude decorrelation, correlators with higher powers of flow harmonics must be considered. Taking as an example  the nonlinear coupling between $V_4$
and $V_2$, the correlation coefficient for the  flow vectors squared  is
\be
\frac{\langle V_2^4  V_4^\star(p)^2\rangle }{\sqrt{{\langle v_2^8  \rangle}{\langle v_4^4(p)  \rangle}}}  \ ,
\ee
the factorization breaking coefficient for the flow magnitudes squared  is
\be
\frac{\langle v_2^4  v_4(p)^2\rangle }{\sqrt{{\langle v_2^8  \rangle}{\langle v_4^4(p)  \rangle}}} \ ,
\ee
and the flow angle correlation can be estimated from the ratio of the above as
\be
\frac{\langle V_2^4  V_4^\star(p)^2\rangle }{\langle v_2^4  v_4(p)^2\rangle } \ .
\label{eq:ang422}
\ee
The correlation coefficient for higher powers of the flow vectors (Fig. \ref{fig:v24v423040}) is smaller  than for the lower powers of the respective
flow vectors (Fig. \ref{fig:v22v43040}). In semi-central collisions, the flow
magnitude decorrelation  (Fig. \ref{fig:mag24v423040}) accounts for about one
half of the flow vector decorrelation shown in Fig. \ref{fig:v24v423040}.

The measure (\ref{eq:ang422}) of  event by event  flow angle correlation for  mixed harmonics involves six-particles correlators and could be  extracted from the  experimental data. We notice that the flow angle correlation  (\ref{eq:ang422}) defined as the ratio of the correlation coefficient for flow vectors and of the factorization breaking  coefficient of the flow magnitudes is a good approximation of  flow angle correlation  weighted with powers of  flow magnitudes
\be
\frac{\langle v_4^2 v_2^4 \cos\left[ 8\left( \Psi_4(p)- \Psi_2 \right)\right] \rangle}{\langle  v_4^2 v_2^4 \rangle } \ .
\label{eq:angmodel422}
\ee
For completeness, in   Fig. \ref{fig:angcentr} we  show the centrality dependence of flow angle decorrelation between momentum averaged flow vectors $V_4^2$ and $V_2^4$. Together with the momentum dependent flow angle correlation shown in Fig. \ref{fig:ang22v43040}, it can serve as an additional experimental observable, sensitive to the model of the initial conditions.

\section{Summary and outlook}

We study observables defined as correlators of several    flow harmonics where one or two of them are measured in a fixed bin of transverse momentum. 
The separate measurement of flow magnitude and flow angle factorization breaking requires the calculation of such four-particle correlators. The  ALICE Collaboration has measured four-particle correlators,
from which the factorization breaking coefficients of the flow vectors squared,  the flow magnitudes squared, 
and  the flow angle correlation can be extracted. Our model simulations indicate  that these measures can be used to estimate the flow angle correlation between  flow harmonics. We find that the  decorrelation  for the flow
vectors squares approximately factorizes  into the flow magnitude
and flow angle decorrelation.

We present calculations of similar correlators for mixed harmonics, e.g. between $V_4(p)$ and $V_2^2$ or between $V_5(p)$ and $V_2V_3$. Such momentum dependent mixed correlations provide additional information sensitive to  the initial state and the dynamics in  the  hydrodynamic evolution.  Correlators for the  squares of the harmonic flow vectors (e.g. between $V_4^2(p)$ and $V_2^4$ or between $V_5^2(p)$ and $V_2^2V_3^2$)
allow to  measure separately the flow vector correlation coefficient, the flow vector magnitudes factorization breaking coefficient, or the flow vector angles correlation.

The simulation results for centrality $30-40$\% do not  reproduce the
preliminary experimental data of the  ALICE Collaboration. Generally,
we find a stronger decorrelation, while in the experiment the harmonic
flow correlation
is sometimes larger than one.
This may indicate that a large contribution of non-flow correlations
is present. This issue requires further study.

An estimate of  correlation measures involving four harmonic flow vectors
can be constructed using separate bins in pseudorapidity, in order to reduce the non-flow correlations. Using four bins in pseudorapidity located at $-\eta_{F}, -\eta, \eta, \eta_{F}$, the factorization coefficient for flow vectors squared can be
estimated as
\begin{widetext}
\be
 r_{n;2}(p)\simeq 
 \frac{\langle V_n(-\eta_{F}) V_n^\star(-\eta,p)V_n^\star(\eta,p) V_n(\eta_{F})  \rangle \langle V_n^\star(-\eta) V_n(\eta) \rangle }{\langle V_n(-\eta_{F}) V_n^\star(-\eta)V_n^\star(\eta) V_n(\eta_{F})  \rangle \langle V_n^\star(-\eta,p) V_n(\eta,p) \rangle } \ , 
\ee
the factorization breaking coefficient between flow vector magnitudes can be defined as
\be
 r_{n;2}^{v_n^2}(p)\simeq 
 \frac{\langle V_n(-\eta_{F}) V_n(-\eta,p)V_n^\star(\eta,p) V_n^\star(\eta_{F})  \rangle \langle V_n^\star(-\eta) V_n(\eta) \rangle }{\langle V_n(-\eta_{F}) V_n(-\eta)V_n^\star(\eta) V_n^\star(\eta_{F})  \rangle \langle V_n^\star(-\eta,p) V_n(\eta,p) \rangle } \ , 
\ee
and the flow angle correlation is estimated from  the ratio the two   quantities
above
\be
F_n(p) \simeq \frac{\langle V_n(-\eta_{F}) V_n^\star(-\eta,p)V_n^\star(\eta,p) V_n(\eta_{F})  \rangle\langle V_n(-\eta_{F}) V_n(-\eta)V_n^\star(\eta) V_n^\star(\eta_{F})  \rangle }{\langle V_n(-\eta_{F}) V_n(-\eta,p)V_n^\star(\eta,p) V_n^\star(\eta_{F})  \rangle\langle V_n(-\eta_{F}) V_n^\star(-\eta)V_n^\star(\eta) V_n(\eta_{F})  \rangle } \ .
 \ee
\end{widetext}
The four-particle correlators in the above formulae involve flow vectors at different pseudorapidity and transverse momentum. Therefore, flow decorrelation  in transverse momentum and in pseudorapidity   combine in the result. However, the decorrelation in the longitudinal direction cancels between the numerator and the denominator, assuming it factorizes from the  decorrelation in transverse momentum. In those formulae we also use the approximation
\be
\frac{\sqrt{\langle v_n^4(p)\rangle}}{\langle v_n^2(p) \rangle } \simeq
\frac{\sqrt{\langle v_n^4\rangle}}{\langle v_n^2 \rangle } \ ,
\ee
as discussed in sect. \ref{sect:results}. It is used in order
to avoid a four-particle correlator where all flow vectors are defined
in a small transverse momentum bins. Note, that the flow vectors at the  forward and backward rapidities $\pm \eta_{F}$ do not require the measurement of particle transverse momenta and forward/backward  calorimeters could used to measure them. Only the flow vectors $V_n(\pm \eta, p)$ require the measurement of the transverse momenta of individual particles. For that purpose two bins  well separated in pseudorapidity in the central rapidity region of the detector acceptance can be used. The separate measurement of the flow magnitudes and the flow angles decorrelation in transverse momentum would provide a sensitive probe to the initial state fluctuations in the heavy-ion dynamics and could further constrain the initial state models.

\vspace{0.5 cm}

\section*{Acknowledgments}

This research is supported by the AGH University of Science and
Technology and  by the  Polish National Science Centre grant
2018/29/B/ST2/00244.

\bibliography{../hydr}

\end{document}